\documentclass[prl,aps,twocolumn,showpacs]{revtex4} 
\usepackage{epsfig}
\date{}

\begin{document}

\title{Modification of scattering lengths  via 
magnetic dipole-dipole interactions}
\author{I.E. Mazets $^{1,2}$ and G. Kurizki $^2$}
\affiliation{$^1$ A.F. Ioffe Physico-Technical Institute, St. 
Petersburg 194021, Russia \\
$^2$ Chemical Physics Dept., Weizmann Institute of Science, 
Rehovot 76100, Israel  } 

\begin{abstract}
We propose a new mechanism for tuning an atomic $s$-wave scattering 
length. The effect is caused by virtual transitions between different  
Zeeman sublevels via magnetic dipole-dipole 
interactions. These transitions give rise to an effective potential, 
which, in contrast to standard magnetic interactions, has an isotropic 
component and thus affects $s$-wave collisions. Our numerical 
analysis shows that for $^{50}$Cr the scattering length can be 
modified up to 15 {\%}. 
\end{abstract} 

\maketitle

Ultracold atomic collisions are essentially $s$-wave collisions.  
The $s$-wave scattering length $a$ is the parameter that 
fully characterizes short-range interactions in an atomic Bose-Einsten 
condensate (BEC) \cite{dalfovo}. Since it is  deisrable to 
control  the interactions in a BEC, tuning the scattering 
length has become an important issue. 
In bulk systems, $a$ can be tuned by means of Feshbach resonance 
\cite{feshbach}. Additionally, low-dimensional systems exhibit 
confinement-induced resonance \cite{cir1d,cir2d}. 

These methods do not affect the  dipole-dipole 
long-range interactions.  It is widely believed 
that short-range and dipole-dipole interactions act {\em independently} 
under usual experimental conditions. 
The influence of static magnetic dipole-dipole interactions on the 
properties of a trapped BEC has been 
studied theoretically \cite{stat-th} and  
experimentally  \cite{stat-ex}. Laser-induced, 
finite-wavelength, dipole-dipole interactions were theoretically 
studied as well \cite{kurizki}. The 
angular dependence of the static dipole-dipole potential is given by 
the second-order Legendre polynomial and thus does not directly affect 
$s$-wave collisions. Its indirect (mediated by the $d$-wave 
channel) influence changes significantly the $s$-wave 
atomic scattering length only if the interacting static electric 
dipole moments are induced by an external electric field of the 
order of 100~kV/cm \cite{ly}, which is a rather high value for an 
ultracold-atom experiment at present. The same  
mechanism is expected to be more efficient for 
highly-polarizable molecules  \cite{shaironen}. 

The conclusion that $s$-wave scattering is hardly affected by 
dipole-dipole interactions has been reached 
under the assumption that the dipole moments 
of the interacting atoms are given by the 
{\em mean values} of the corresponding 
operators, while transitions to other Zeeman sublevels have been neglected. 
In the present Letter we demonstrate that these 
(virtual) transitions may lead to an appreciable change in the value of $a$. 
In this context, we discuss a remarkable general effect that has been 
overlooked thus far: the {\em nonlocal} 
collisional dynamics of ultracold atoms coupled by interaction between 
{\em fluctuating} magnetic dipoles. 

Since the magnetic moment of an atom in a weak magnetic field is given 
by $g{\hat{\bf F}}$, $g$ being the gyromagnetic factor and $\hat{\bf F}$ 
the total angular momentum, the dipole-dipole 
interaction operator is 
\begin{equation} 
\hat{V}_{DD}({\bf r})=\frac{ \mu _0g^2}{4\pi } 
\frac{\hat{\bf F}_1 \hat{\bf F}_2-3 
(\hat{\bf F}_1 {\bf e}_r)(\hat{\bf F}_2{\bf e}_r)}{r^3}, 
\label{eq1} 
\end{equation} 
where {\bf r} is the interatomic separation vector, ${\bf e}_r={\bf r}/r$. 
Suppose the degeneracy of states with different $z$-projection of 
the momentum, $F_z\equiv M$, is lifted by 
an external magnetic field. Let the atoms be in the lowest 
energy state $|F, \, M=-F\rangle $. The excitation energy for the 
Zeeman sublevel $|F, \, M=-F+1\rangle $ is denoted by $\Delta E$. 
The operator 
(\ref{eq1}) couples the two-atom spin state 
\begin{equation}
|\alpha \rangle =
|F, M=-F\rangle _1  |F, M=-F\rangle _2 
\label{a1a}
\end{equation} 
to 
\begin{eqnarray} 
|\beta \rangle &=&(1/\sqrt{2})(   
|F, M=-F\rangle _1  |F, M=-F+1\rangle _2 \nonumber \\ && 
+|F, M=-F+1\rangle _1  |F, M=-F\rangle _2 ).     \label{a1b} 
\end{eqnarray} 
We adopt the two-channel 
approximation by representing the wave functions of the colliding pair 
as $|\psi \rangle =\psi _\alpha ({\bf r})|\alpha \rangle +\psi _\beta 
({\bf r})|\beta \rangle $. Then the Schr\"odinger equation reads as 
\begin{eqnarray} 
\frac {\hbar ^2k^2}m\psi _\alpha ({\bf r})&=&-\frac {\hbar ^2}m \nabla ^2 
\psi _\alpha ({\bf r})+\langle \alpha |V_{DD}({\bf r})|\alpha 
\rangle \psi _\alpha ({\bf r})\nonumber \\ &&
+\langle \alpha |V_{DD}({\bf r})|\beta \rangle \psi _\beta ({\bf r}), 
\label{eq2a} \\
-\frac {\hbar ^2q^2}m\psi _\beta ({\bf r})&=&-\frac {\hbar ^2}m \nabla ^2 
\psi _\beta ({\bf r})+\langle \beta |V_{DD}({\bf r})|\beta 
\rangle \psi _\beta ({\bf r})\nonumber \\ &&
+\langle \beta |V_{DD}({\bf r})|\alpha \rangle \psi _\alpha ({\bf r}), 
\label{eq2b}
\end{eqnarray} 
Here $m$ is the atomic mass, $\hbar ^2k^2/m$ is the collision energy, and 
$\hbar ^2q^2=\Delta E -\hbar^2k^2/m$. Since we consider ultracold 
collisions, we have 
\begin{equation} 
\hbar ^2q^2/m \approx \Delta E. \label{a2}
\end{equation}  

The next step is to assume that $\Delta E$ is large enough to allow us 
to neglect the term $\langle \beta |V_{DD}({\bf r})|\beta \rangle \psi _
\beta ({\bf r})$ in Eq.~(\ref{eq2b}). 
Then the virtually excited state $|\beta 
\rangle $ can be eliminated using the negative-energy Green function of 
the free motion $G(-\Delta E,{\bf r})=-e^{-q|{\bf r}|}/(4\pi |{\bf r}|)$ 
\cite{sakurai}: 
\begin{equation} 
\psi _\beta ({\bf r}) =\int d^3{\bf r}^\prime \, G(-\Delta E, {\bf r}-
{\bf r}^\prime )\langle \beta |V_{DD}({\bf r}^\prime )|\alpha \rangle 
\psi _\alpha ({\bf r}^\prime ) .
\label{eq3}
\end{equation} 
Substituting Eq. (\ref{eq3}) into Eq. (\ref{eq2a}), we obtain a 
{\em non-local}  equation \cite{nakamura} for $\psi _\alpha $. 
\begin{eqnarray} 
\frac {\hbar ^2k^2}m\psi _\alpha ({\bf r})&=&-\frac {\hbar ^2}m \nabla ^2 
\psi _\alpha ({\bf r})+\langle \alpha |V_{DD}({\bf r})|\alpha 
\rangle \psi _\alpha ({\bf r})\nonumber \\ &&
+\langle \alpha |V_{DD}({\bf r})|\beta \rangle \int d^3{\bf r}^\prime \, 
G(-\Delta E,{\bf r}-{\bf r}^\prime ) \nonumber \\ && 
\times \langle \beta |V_{DD}({\bf r^\prime })
|\alpha \rangle \psi _\alpha ({\bf r}^\prime ).
\label{eq4}
\end{eqnarray} 
On the assumption (verifiable {\em a posteriori}) that both $G$ and 
$\langle \beta |V_{DD}|\alpha  \rangle $ vary faster than $\psi _\alpha $ 
with ${\bf r}^\prime $, we can pull the wave function out of the integral 
and calculate the necessary matrix elements on expanding  
$G(-\Delta E,{\bf r}-{\bf r}^\prime ) $ as a 
series in spherical harmonics \cite{varsh}. 
The resulting Schr\"odinger equation for a pair of 
colliding atoms with maximal projection of the angular momentum still 
{\em retains the signature of nonlocal dynamics}, by virtue of the 
spatial dependence of the effective potential on the convolution of 
the Green function  $G(-\Delta E,{\bf r}-{\bf r}^\prime ) $ 
and the matrix element  $\langle \beta |V_{DD}({\bf r}^\prime  ) 
|\alpha  \rangle $. It has the form  
 
\begin{eqnarray} 
\frac {\hbar ^2k^2}m\psi _\alpha ({\bf r})&=&-\frac {\hbar ^2}m \nabla ^2 
\psi _\alpha ({\bf r}) \nonumber     \\ &&
+\frac {\mu _0g^2F^2}{4\pi r^3}(1-3\cos ^2 \theta ) 
\psi _\alpha ({\bf r}) \label{eq5}   \\ &&
-\frac {9\mu _0^2mqg^4F^4}{16 \pi ^2\hbar ^2 r^3} 
\cos ^2\theta \sin ^2 \theta \, W(qr) 
\psi _\alpha ({\bf r} ), 
\nonumber 
\end{eqnarray} 
where $\theta $ is the angle between {\bf r} and the quantization axis 
and 
\begin{widetext} 
\begin{eqnarray} 
W(qr)&=&\int _0^{qr}\frac {d\xi }\xi \, \left[ \left(
\frac 3{\xi ^3}+\frac 1{\xi }\right) \sinh \xi  
-\frac 3{\xi ^2 }\cosh \xi  \right]  
\exp(-qr)[3(qr)^{-3}+3(qr)^{-2}+(qr)^{-1} ] \nonumber \\ &&
+\int _{qr}^\infty \frac {d\xi }\xi \, \exp(-\xi )(3\xi ^{-3}+
3\xi ^{-2}+\xi ^{-1}) \left[ \left(
\frac 3{q^3r^3}+\frac 1{qr }\right) \sinh qr  
-\frac 3{q^2r^2 }\cosh qr  \right] .
\label{eq6} 
\end{eqnarray} 
\end{widetext} 
The function $W(qr)$ is plotted in Fig. 1. It has two asymptotes:  
\begin{equation} 
W(qr)\approx (qr)^{-3}, \qquad qr \gg 1,       \label{a3a} 
\end{equation} 
which corresponds to 
adiabatic elimination of $\psi _\beta $ in the limit wherein the 
approximation $G(-\Delta E,{\bf r}-{\bf r}^\prime )\approx - q^2
\delta ({\bf r}-{\bf r}^\prime )$ holds (dashed line in Fig.~1), and  
\begin{equation} 
W(qr)\approx (6qr)^{-1},   \qquad qr\ll 1     \label{a3b} 
\end{equation}  
(dot-dashed line in Fig.~1). 

\begin{figure} 
\begin{center} 
\centerline{\epsfig{file=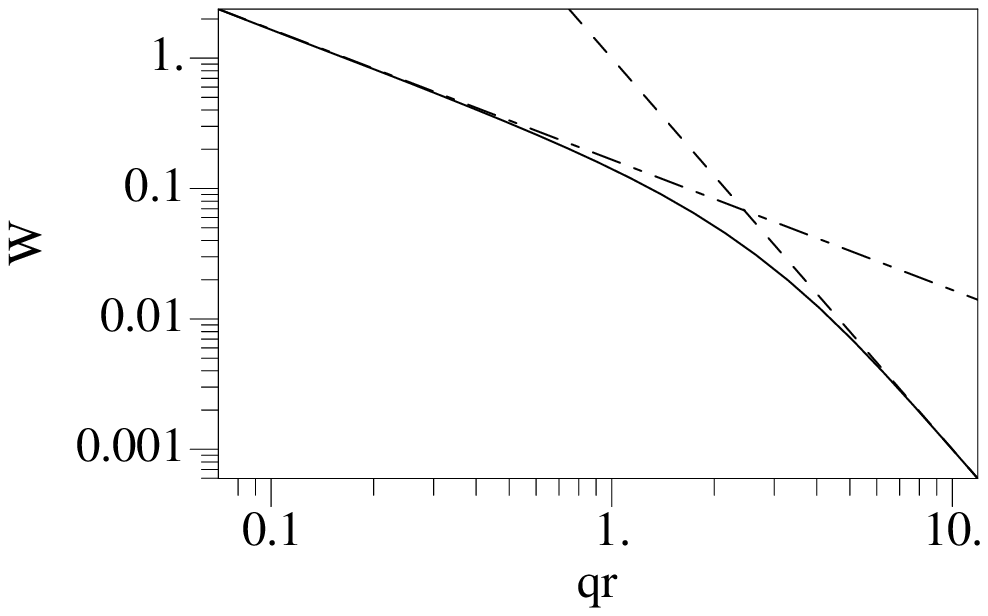,width=7cm}}
\end{center}

\begin{caption} 
{Dimensionless function $W$ vs. dimensionless distance $qr$. The two 
power-law asymptotes are shown (see details in the text).} 
\end{caption} 
\end{figure} 

To analyze the influence of the additional potential mediated by 
static magnetic dipole-dipole interactions, we average Eq.~(\ref{eq6}) over 
all angles and obtain in the ultracold-collision limit ($k\rightarrow 0$) 
the following equation: 
\begin{equation} 
\frac 1{r^2} \frac d{dr} \left( r^2 \frac d{dr} \psi _\alpha \right) +
\frac {9ql_m^2W(qr)}{8r^3}\psi _\alpha =0,  
\label{eq7a} 
\end{equation}
where the quantity 
\begin{equation} 
l_m = \mu _0mg^2F^2 /(4 \pi \hbar ^{2})    
\label{eq7b} 
\end{equation} 
is the characteristic magnetic length. Equation (\ref{eq7a}) 
must be supplemented by the hard-sphere boundary condition 
\begin{equation} 
\psi _\alpha (a) =0 
\label{eq8} 
\end{equation} 
that accounts for the short-range (van der Waals) interactions. 
Solution of Eqs.~(\ref{eq7a}), (\ref{eq8}) yields the modified scattering 
length $\tilde{a}$, which can be inferred from the $r\rightarrow \infty $ 
asymptotics of the solution that is propoportional to $1-\tilde{a}/r$. 

\begin{figure} 
\begin{center}
\centerline{\epsfig{file=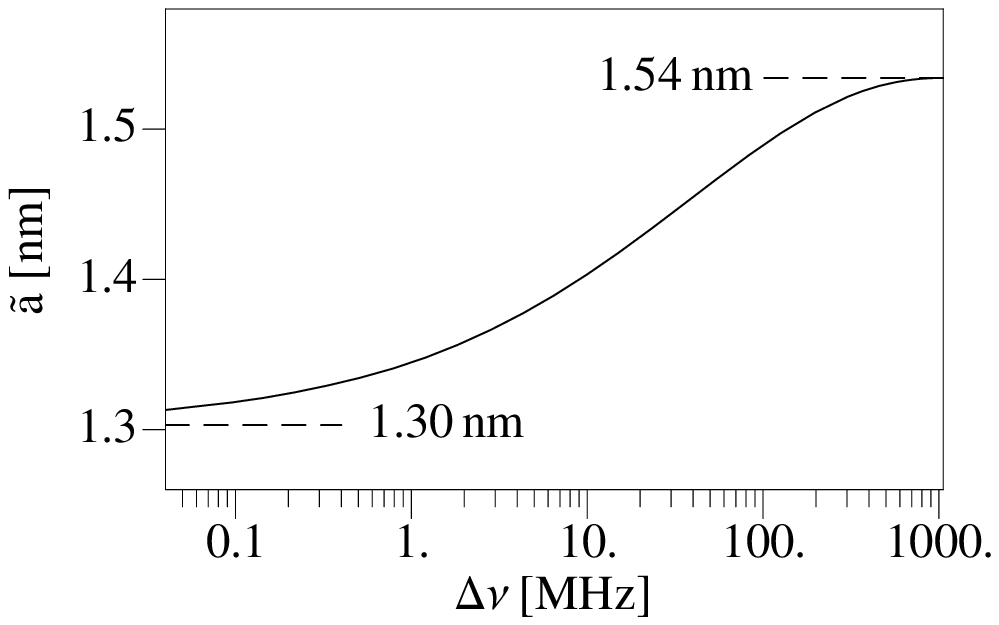,width=7cm}}
\end{center} 

\begin{caption} 
{Modified scattering length for $^{50}$Cr vs.  frequency splitting 
$\Delta \nu =\Delta E/(2\pi \hbar )$. The two asymptotes of 
$\tilde{a}$ in the small- and large-$q$ limits are shown.  }
\end{caption} 
\end{figure}

It is noteworthy that for $q\rightarrow 0$ the magnetically-induced 
isotropic potential varies as $r^{-4}$ and does not depend on $q$. In 
this case  Eqs.~(\ref{eq7a}), (\ref{eq8}) admit an analytical solution, 
\begin{equation}
\psi _\alpha (r)=\sin \left[ \frac{\sqrt{3}l_m}{4a}
\left( 1 -\frac ar\right)  \right] , 
\label{a4} 
\end{equation}  
so that the modified scattering length becomes 
\begin{equation} 
\tilde{a}\vert _{q\rightarrow 0} =a \frac {\sqrt{3}l_m/(4a)}{\tan 
[\sqrt{3}l_m/(4a)]}. 
\label{eq9} 
\end{equation} 

In Fig. 2 we plot the results of numerical integration of Eq. (\ref{eq7a}) 
for $^{50}$Cr. This isotope is chosen because chromium atoms have 
quite a large magnetic moment, $gF = 6\mu _B$, $\mu _B$ being the Bohr 
magneton. On the other hand, the bare (unaffected by the effect under 
consideration) scattering length $a$ for this isotope in the state of 
the maximum spin projection is of about 
29 Bohr's radii, as can be concluded by mass-scaling arguments from the 
most recent data on $^{52}$Cr scattering \cite{crscat}. Note, that the 
recent, more accurate measurements \cite{cranew} 
yield smaller values of the scattering length for $^{52}$Cr and, 
consequently, for $^{50}$Cr, than the previously reported data 
\cite{craold}. The value 
$a=1.54$~nm is smaller than $l_m=2.34$~nm, and the difference between 
$a$ and $\tilde{a}$ becomes observable (up to 15 \%). By contrast, 
the scattering length for $^{52}$Cr is three times larger than for 
$^{50}$Cr. As a result, the influence of static magnetic dipole 
interactions in the case of $^{52}$Cr is negligible. 

The discussed mechanism of the scattering length modification 
will manifest itself via excitation of BEC oscillations by periodic 
modulation of the external uniform magnetic field. Such a method to 
reveal the scattering length modulation has been proposed with regard 
to Feshbach resonance \cite{kagan}. Consider a BEC of $^{50}$Cr 
in the Thomas-Fermi regime \cite{pethick} trapped in an optical dipole 
harmonic potential, which is assumed, for the sake of simplicity, to 
be isotropic, $\omega _0$ being its fundamental frequency. The uniform, 
$z$-directed magnetic field oscillates in time as 
\begin{equation}
B(t)=B_0+B_1 \sin \omega t.               \label{a5} 
\end{equation}  
Its variation results in variation of the frequency 
splitting $\Delta \nu $ between the states $|\alpha \rangle $ and 
$|\beta \rangle $. The Zeeman shift in a weak field is given by 
$\Delta \nu /B =87.7$~MHz/mT. Thus we can infer the time variation  
of the scattering length using the dependence $\tilde{a}(\Delta \nu )$ 
plotted in Fig.~2. 

Let the magnetic field oscillations be switched on  at $t=0$. 
The lowest monopole (angle-independent) mode of the BEC oscillations 
is described by the nonlinear equation \cite{kagan} 
\begin{equation} 
\ddot{b}+\omega _0^2b=\tilde{a}(t)/(\tilde{a}_0b^4), 
\label{a6}
\end{equation} 
where $b=R/R_0$ is the ratio of the Thomas-Fermi radius $R$
of the oscillating BEC to the BEC radius at rest, $R_0= 
R\vert _{t=0}=l_{ho}(N\tilde{a}_0/l_{ho})^{1/5}$ \cite{pethick},  
$\tilde{a}_0= \tilde{a}\vert _{t=0}$, $l_{ho}=\sqrt{\hbar 
/(m\omega _0)}$, $N$ is the number of atoms in the BEC. For the 
numerical estimations we take $\omega _0/(2\pi )\sim 100$~Hz, 
$N\sim 10^5$, thus obtaining $R_0\sim 6~\mu $m and the peak 
density of the BEC of about $3\cdot 10^{14}$~cm$^{-3}$. 
The initial conditions to Eq. (\ref{a6}) are 
$b\vert _{t=0}=1$, $\dot{b}\vert _{t=0}=0$. 

\begin{figure} 
\begin{center} 
\centerline{\epsfig{file=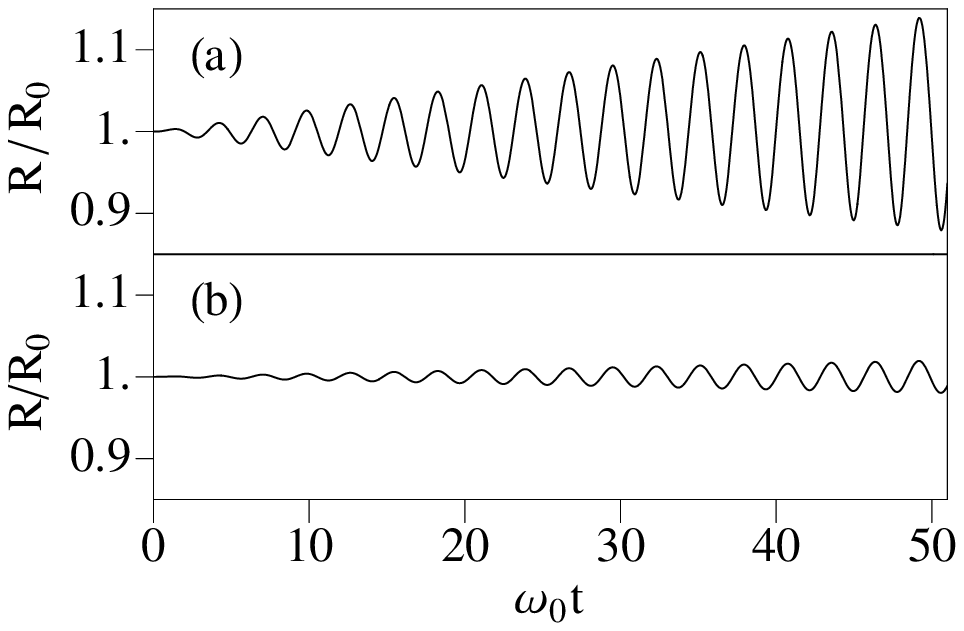,width=7cm}}
\end{center} 

\begin{caption}
{Relative (dimensionless) modulation of the Thomas-Fermi BEC 
radius vs. dimensionless time under the action of the variable 
magnetic field Eq.~(\ref{a5}). (a) $B_0=0.08$~mT, (b) $B_0=0.6$~mT; 
$B_1=0.04$~mT in both the cases. }   
\end{caption} 
\end{figure}

Choosing the frequency of the magnetic field modulation 
$\omega =\sqrt{5}\omega _0$, we ensure the resonance 
of the parametric excitation with the lowest monopole mode in the 
linear regime. In Fig.~3 we present the numerical solution of 
Eq.~(\ref{a6}) for two different choices of the magnetic field 
parameters. While the amplitude $B_1$ 
of the magnetic field oscillations is the same, the mean value, $B_0$, 
takes two distinct values. In Fig.~3(a) $B_0$ corresponds to the 
steepest derivative $d\tilde{a}/dB$, so that the BEC excitations 
are excited most efficiently. For our 
numerical estimations we take $\omega _0/(2\pi )\sim 100$~Hz, 
$N\sim 10^5$, thereby obtaining the Thomas-Fermi BEC radius 
$R_0\sim 6~\mu $m and the peak 
density of the BEC of about $3\cdot 10^{14}$~cm$^{-3}$. Thus 
for the parameters of Fig.~3(a), after 80~ms the amplitude of 
oscillations of the Thomas-Fermi BEC radius approaches $0.84~\mu $m. This  
amplitude is approximately twice larger than the resonant wavelength 
of the optical $^7S_3\leftrightarrow ^7P_3$ transition in chromium, 
thus making the BEC oscillations detectable by optical imaging. 
By contrast, the value of $B_0$ in Fig.~3(b) corresponds to a 
relatively small $d\tilde{a}/dB$ and, therefore, to much less 
efficient excitation of the BEC.  

To conclude, we have proposed a new mechanism of $s$-wave scattering 
length tuning and modulation,  
different from both Feshbach-resonance \cite{feshbach} and 
confinement-induced resonance methods \cite{cir1d,cir2d}. It involves 
a hitherto unexplored aspect of ultracold-atom collisions, namely, 
their nonlocality mediated by fluctuations of the magnetic dipole 
moments. Magnetic dipole-dipole interaction causes  
virtual transitions between different 
Zeeman sublevels of the atomic ground state. 
Such a coupling to the virtual state induces, additionally to van 
der Waals forces, an attractive {\em isotropic} 
interatomic potential. If the energy gap $\Delta E$ between 
the occupied and virtual spin states grows to infinity, we 
recover the ``bare'' (unmodified) scattering length value $a$, which is 
calculated from the van der Waals (non-magnetic) potential only. 
As $\Delta E$ vanishes, the renormalized scattering length approaches 
the limit (\ref{eq9}) that is fully determined by the ratio of the 
magnetic length $l_m$, Eq. (\ref{eq7b}), to $a$. An appreciable effect 
is expected for $l_m/a \gtrsim 1$, which is the case for $^{50}$Cr 
atoms, where we expect lowering of the modified scattering length by 
up to 15 \% with respect to $a$.   

This work is partially supported by the German-Israeli Foundation, 
the EC (SCALA NOE), and ISF. I.E.M. also acknowledges 
support from the program Russian Leading Scientific Schools (grant 
9879.2006.2). The authors thank Prof. T. Pfau for helpful discussions.

\end{document}